\documentclass[twocolumn,prl,aps]{revtex4}
\usepackage{amsmath}
\usepackage{amssymb}
\usepackage{graphicx}
\usepackage{dcolumn}
\usepackage{bm}
\usepackage{xcolor}
\usepackage{epstopdf}

\begin{document}

\title{Transport through the network of topological channels in HgTe based quantum well}
\author{G. M. Gusev, $^1$  Z. D. Kvon, $^{2,3}$ D. A. Kozlov, $^2$  E. B. Olshanetsky, $^{2}$ M. V. Entin, $^{2}$ and   N. N. Mikhailov $^{2,3}$}

\affiliation{$^1$Instituto de F\'{\i}sica da Universidade de S\~ao
Paulo, 135960-170, S\~ao Paulo, SP, Brazil}
\affiliation{$^2$Institute of Semiconductor Physics, Novosibirsk
630090, Russia}
\affiliation{$^3$Novosibirsk State University, Novosibirsk 630090,
Russia}

\date{\today}
\begin{abstract}

Topological insulators represent a new quantum state of matter which is characterized by edge or surface states and an insulating band gap in the bulk. In a two dimensional (2D) system based on the HgTe quantum well (QW) of critical width random deviations of the well width from its average value result in local crossovers from zero gap 2D Dirac fermion system to either the 2D topological insulator or the ordinary insulator, forming a complicated in-plane network of helical channels along the zero-gap lines. We have studied experimentally the transport properties of the critical width HgTe quantum wells near the Dirac point, where the conductance is determined by a percolation along the zero-gap lines. The experimental results confirm the presence of percolating conducting channels of a finite width. Our work establishes the critical width HgTe QW as a promising platform for the study of the interplay between topology and localization.

\end{abstract}

\maketitle

\section{Introduction}
Topological states of matter have attracted a lot of attention due to their numerous intriguing transport properties.  In particular, in two-dimensional topological insulators (2D TI)
there are gapless conducting helical edge channels, that are protected against backscattering \cite{kane, hasan, qi, moore, moore2}. Many experiments have been performed to investigate
the transport properties of the 2DTI edge states in several materials where the helical edge states are due to different physical mechanisms. For example the main factor responsible for
the existence of a nontrivial topological phase in HgTe/CdTe quantum wells (QW) is a strong spin–orbit interaction \cite{konig, buhman, roth, gusev, olshanetsky, rahim}, whereas in InAs/GaSb
double QWs the topologically protected helical edge states in the inverted phase emerge as a consequence of the coupling between the bands leading to the opening of a hybridization
gap \cite{knez, knez2, du, nichele, suzuki}. As has been verified on many occasions (for review see \cite{bernevig}) generally two basic experimentally observed features indicate
the presence of ballistic helical edge channels in submicron 2DTI samples: the conductance of the order of the universal value $e^{2}/h$ \cite{konig, olshanetsky} and a strong
nonlocal signal due to the helical edge states current circulating along the sample perimeter  \cite{roth, gusev, olshanetsky, rahim}.

\begin{figure}[ht]
\includegraphics[width=\linewidth]{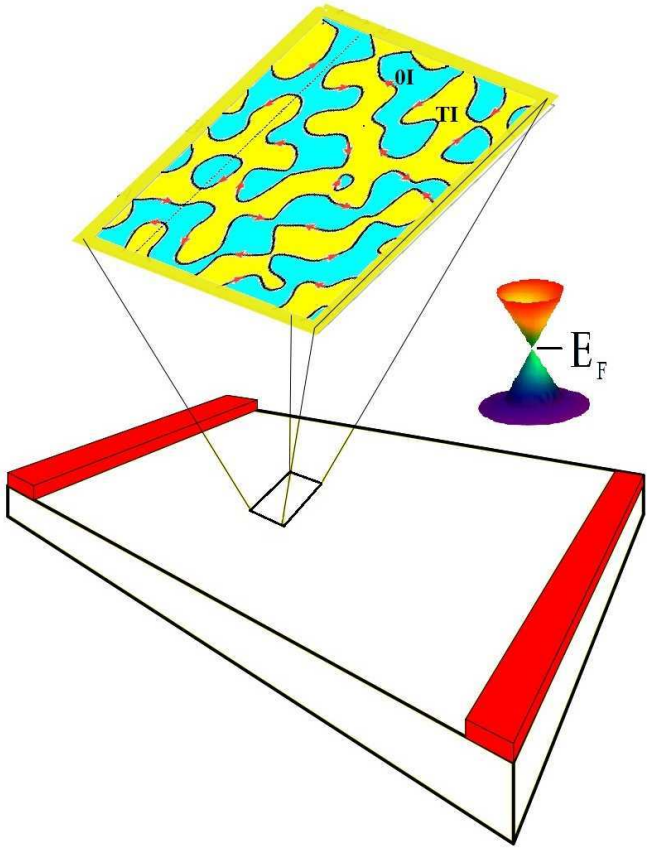}
\caption{(Color online)
Schematic of zero energy topological channels in the slab shaped sample based on HgTe quantum well of a critical width $w_{c}\approx 6.3 nm$.
Yellow regions are topological insulator domains, blue regions correspond to ordinary insulator domains. The Dirac Fermions motion direction at
a fixed spin projection is shown by arrows. }
\end{figure}

The interplay between topology and localization is another challenging object of study both for theoreticians and experimentalists. To gain a deeper insight into the
critical behaviour of matter, theoreticians often employ network models. Such is the case for the metal-insulator transition and also for the transition between
different phases of topological insulator \cite{onoda, obuse, obuse2, bondesan, yamakage, bhardwaj}. Critical phenomena related to the integer and fractional quantum
Hall effects have been successfully described by a chiral Chalker-Coddington-like network representation of bulk transport in the high magnetic field limit \cite{chalker}.
 In contrast to the quantum Hall states, the two-dimensional topological insulator-metal transition  could be represented by uncoupled counter propagating channels with
 opposite spin, the so called $Z_{2}$ network model \cite{onoda, obuse, bondesan,yamakage, bhardwaj}.

It was recognized that for topological metallic states a well-defined mobility edge, i.e. a specific energy separating the region of extended states
from that of the localized states, is expected, while the extended states in quantum Hall effect are located at particular distinct energy values ( no mobility edge).

As already mentioned, a network of conducting channels  occurs naturally in HgTe quantum wells of critical width $w_{c}\approx 6.3 nm$. It has been shown \cite{konig, buhman} that
if the HgTe QW width is below the critical value, the system is an ordinary insulator with a normal energy band structure. If, on the other hand, the QW width is above the critical,
then it is a 2D topological insulator with an inverted energy spectrum. Finally, the QW width $w = w_{c}$ corresponds to the 2D Dirac fermion system with a gapless, single-cone
spectrum (Figure  1). In-plane fluctuations of the QW width about its average value $w = w_{c}$, that cannot be completely avoided during the QW growth, lead to spacial
gap variations (random gap sign changes), and therefore, to transitions between the mentioned topological phases. This results in a network of zero energy channels running along
the boundaries separating the normal insulator and the 2D topological insulator phases (Figure 1). In addition to the gap fluctuations there will also be variations
of the electrostatic potential due to random distribution of charged impurities. For the Fermi energy located near the Dirac point the system conductivity is attributed to the percolation along zero energy channels \cite{mahmoodian, mahmoodian2}.

In the present paper we study the transport properties of the zero gap HgTe quantum well (QW of critical width) both for $B=0$ and in the presence of magnetic field. We
find that the conductivity  of the samples lies in the interval $(1.5-4)e^{2}/h$, as is expected for the topological network formed by finite width conducting channels.
In order to estimate the width of the channels we performed additional measurements of the Hall effect at low magnetic fields. Surprisingly, we find that the Hall effect
near the percolation threshold is completely suppressed in a narrow interval of carrier densities in agreement with the percolation model \cite{mahmoodian, mahmoodian2}.

\section{RESULTS AND DISCUSSION}

\subsection{Structural properties}
Quantum wells $Cd_{0.65}Hg_{0.35}Te/HgTe/Cd_{0.65}Hg_{0.35}Te$ with (013) surface orientations and a nominal well thickness of (6.3-6.6) nm were prepared by molecular beam
epitaxy ( see Figures 2,3). Transmission electron microscopy (TEM) of cross sectional sample allows to
study the interfaces of layered structures directly with  a spatial resolution
down to an nanometer scale.
Figure 2a shows TEM images of the cross sections of the specimen with designations  of various layers constituting the structures, which allow
assessment of abruptness of interfaces as well as lateral uniformity of layer growth \cite{yahniuk}.  The difference of contrasts in TEM images is due to the difference in the chemical compositions of the layers,
which allows to determine the fluctuation of the width of the individual layers in the image of a multilayer structure. Both images indicate that
HgTe/HgCdTe  interfaces are reasonably abrupt. A histogram in Figure 2b is a  display of statistical information on HgTe well width fluctuations.
The histogram  follows a normal Gaussian distribution with mean with $d_{c}\approx 6.1 nm$ with standard deviation of 0.6 nm. As we mentioned above, the width fluctuations near topological transition lead to spacial
gap variations, and therefore, to transitions between the topological phases.
\begin{figure}[ht]
\includegraphics[width=\linewidth]{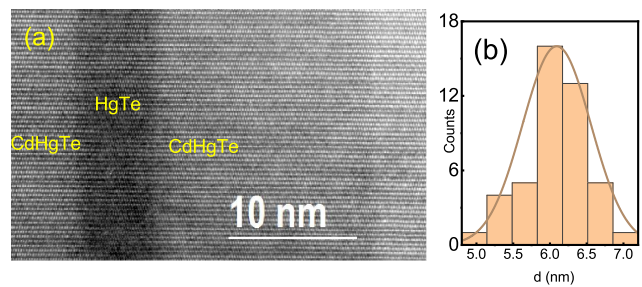}
\caption{(Color online)
(a) TEM images of cross sections of sample. (b) Histogram  displaying the distribution of well widths
across the whole image.}
\end{figure}

\begin{table}[ht]
\centering
\begin{tabular}{|l|l|l|l||l||l||l|}
\hline
sample & $w$ (nm) & $V_{CNP}$ (V)& $\rho_{max}(h/e^{2})$& $\mu (V/cm^{2}s )$ \\
\hline
1&    6,3 & -0.6& 0.35 & 56.000\\
\hline
2&   6.4 & -1.28 & 0.27 &90.000\\
\hline
3&   6.3 & -3 & 0.27 &59.600 \\
\hline
4&   6.3 & -4.3& 0.31 &58.600 \\
\hline
5&   6.3 & -4.7& 0.29 &46.400 \\
\hline
\end{tabular}
\caption{ Some of the typical parameters of the electron system in HgTe quantum well at T=4.2K.}
\end{table}

\subsection{Gapless Dirac fermions and Drude conductivity}

Recently it has been demonstrated that HgTe quantum well with a critical width $w_{c}=6.3 nm$ constitutes a system of 2D fermions with a single Dirac cone spectrum \cite{buttner, kvon, kvon2, gusev2}.
The fluctuations of the QW width result in a random potential in the bulk of the QW. In QW of critical width these fluctuations lead to the formation of a random network of zero energy lines,
as shown in Figure 1. We believe that the physical properties of this network are described by $Z_{2}$ quantum network model, and below we present experimental evidence that supports this assumption.
\begin{figure}[ht]
\includegraphics[width=\linewidth]{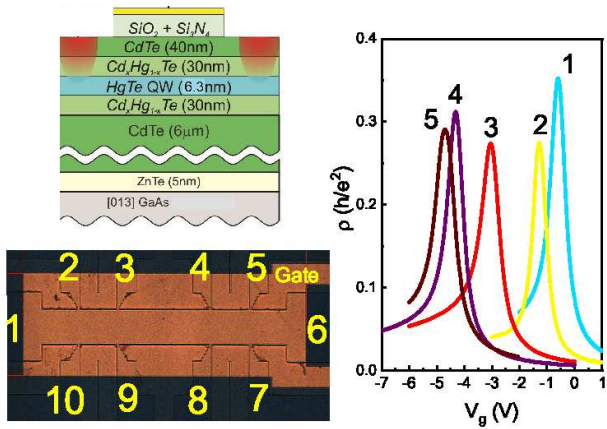}
\caption{(Color online)
Resistance as a function of the gate voltage for different samples, T=4.2K. Left bottom- schematic structure of the sample. Right bottom - top
view of the sample. }
\end{figure}

 \begin{figure}[ht]
\includegraphics[width=\linewidth]{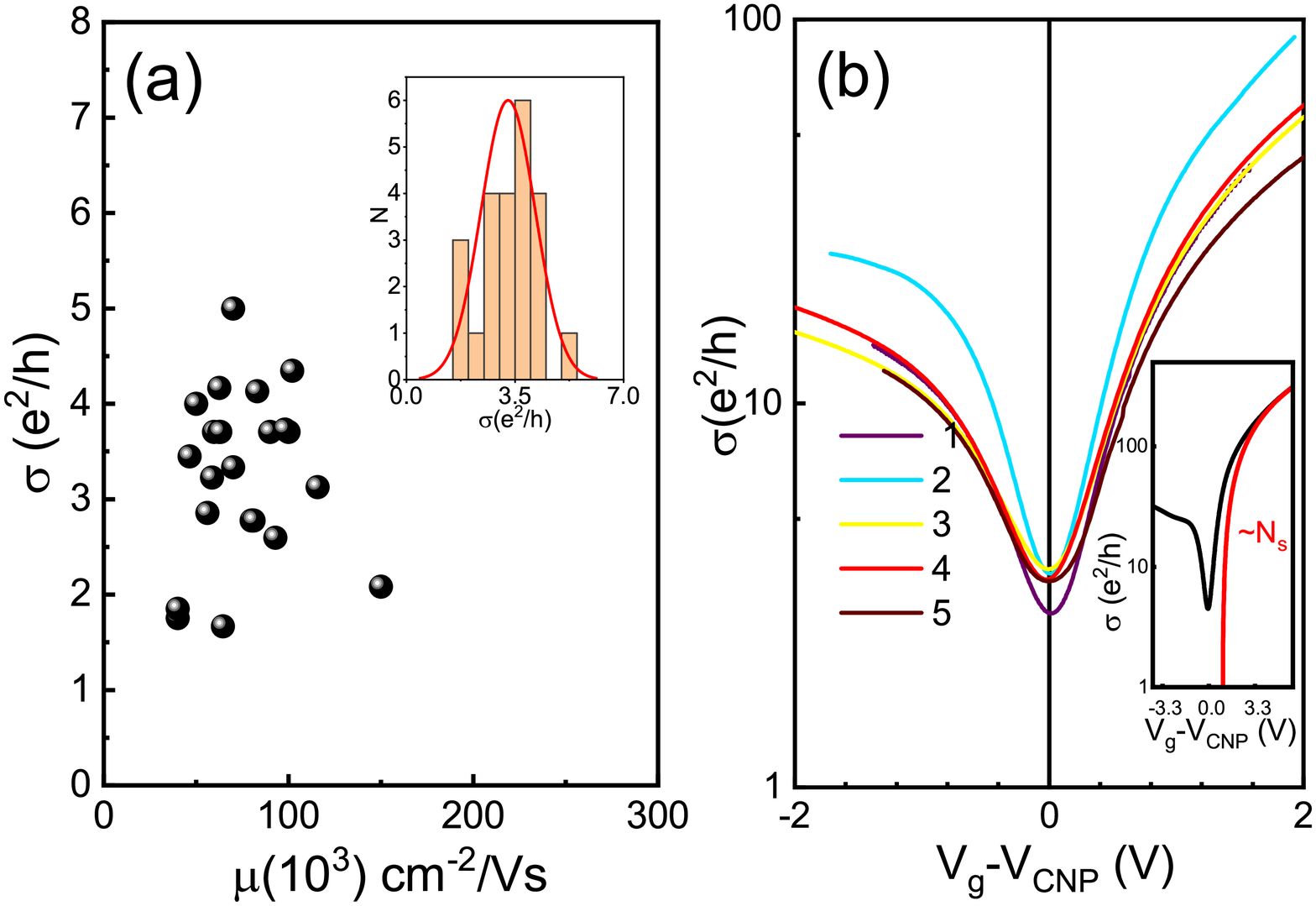}
\caption{(Color online)(a) The minimum conductivity and mobility at $n_{s}=10^{11} cm^{-2}$ for 24 samples. Insert- Histogram  displaying the distribution of the conductivity minimum.
(b) Conductivity of the five representative samples near the CNP as a function of $V_{g}$. Insert- conductivity of sample 1 in a wider range of the gate voltages. Red line-linear $n_{s}$ dependence.}\label{fig:4}
\end{figure}

Devices for transport measurements ( see chapter method) are  specially designed for multi-terminal measurements and consists of three narrow (  $50 \mu m$ wide) consecutive segments of different length
(100, 250, 100 $\mu m$ )and seven voltage probes (see Figure 3, left bottom panel). A dielectric layer was deposited (100 nm of $SiO_{2}$ and 100 nm of $Si_{3}Ni_{4}$) on the sample surface and then covered by a TiAu gate (see Figure 3, left top panel).
In conventional transport  measurements the current is applied between contacts 1–6 and the potential difference is measured
between contacts 2–3, 3–4 and 4–5 of the sample.
Figure 3 shows the resistivity  $\rho(Vg)$ at zero magnetic field for five samples fabricated from different wafers and at different time. The table 1 lists the typical
parameters of the devices, such as the well width $w$, the gate voltage corresponding to the Dirac point position $V_{CNP}$ , the resistivity value at the CNP $\rho_{max}$ and the electron
mobility $\mu = \sigma/n_{s}e$ for the density $n_{s} = 10^{11}cm^{-2}$. The  Figure 4a shows the conductivity distribution for 24 samples, grown during a five years period. One can
see that the conductivity values lie in the interval $(1.5  - 4)e^{2}/h$. Figure 4b shows the conductivity as a function of gate voltage $V_g$ for five representative samples with
parameters listed in table 1. The $\sigma(V_g)$ dependence is symmetric and nearly parabolic close to the CNP for low  electron densities $(n_{s} < 2\times10^{10} cm^{-2})$ and
approximately linear in electron density for higher density values. It is convenient to consider the total conductivity as a sum of the network conductivity $\sigma_{nw}$ and the bulk
conductivity $\sigma_{2D}$: $\sigma_{tot}  = \sigma_{nw} + \sigma_{2D}$.

Let us start our analysis with the bulk contribution to the conductivity. The unscreened Coulomb disorder induced by randomly distributed charge impurities has been considered as the dominant
mechanism of scattering in HgTe quantum well because of a very large HgTe dielectric constant and small Dirac fermions effective mass. Two distinct transport regimes can be indicated. For
large carrier density the Boltzmann transport theory is valid \cite{kvon3}. On approaching the Dirac point (low carrier density), at a certain energy there will be an Anderson transition
from the higher energy delocalized states to the localized states in narrow impurity bands located in small direct or inverted energy gaps (see Figure 1). It would be reasonable
to suppose that transport via the delocalized states near the boundary with the localized bands can be described using the ordinary expression for mobility of carriers with a parabolic
spectrum in the presence of impurity scattering. In this case the mobility is given by $\mu = A(n_{s}/n_{i})$, where $n_{i}$ is the impurity density,  $A=(\varepsilon_{s}/e^{3})(16\pi \hbar^{3}/m_{DF}^{2})$, $m_{DF}$
is the effective mass at the Fermi level \cite{davies}. We get $\sigma_{2D} = e\mu n_{s}$  with quadratic dependence on electron density  for $n_{s} < 2\times 10^{10} cm^{-2}$, as shown in Figure 4b.
 At higher electron densities the electron energy spectrum becomes linear and one has to use the transport time expression calculated in \cite{kvon3}, which leads to a conductivity $\sigma(n_{s})$ nearly
 linear with the electron density. Such conductivity behaviour is shown in Figure 4b.

\subsection{Comparison with the network model}

Transport in HgTe quantum wells of a critical width $w_{0}$ is believed to be governed by the energy gap fluctuations leading to the formation of the topological channels network (Figure 1).
One can assume that the QW width fluctuations $\delta w$ can be described by the Gaussian distribution around the medium width value $w$, which follows from TEM images of the sample (Figure 2).

The percolation network can be characterized by a dimensionless parameter $\xi=\mbox{erf}((w-w_0)/\sqrt{2}\delta w)$, where $\mbox{erf}(x)$ is the complementary error function. The unavoidable QW width
fluctuations lead to a sample breaking-up into domains with positive (the ordinary insulator (OI)) and negative (the topological insulator (TI)) gap signs. If $|w-w_0|\gg\delta w$ and $w>w_0$ ($\xi\to 1$),
then rare OI domains are embedded in TI domains. If, on the other hand,  $w<w_0$, $\xi\to 1$, then rare TI domains are embedded in OI domains. For $w\to w_0$ ($\xi\to \xi_c=0.5$) these domains are mixed in
approximately equal proportion.

At the CNP the low-temperature the bulk conductivity of the OI or the TI domains should be close to zero. However, in the immediate vicinity of the lines separating the OI and TI domains (the zero gap line
(ZGL)) the QW is a nearly gapless conductor. In fact, any such ZGL can be considered as a quantum wire. The sample conductance will be non-zero if such a line runs from one end of the sample to the other.
On approaching the percolation threshold there forms a whole network of ZGLs covering the entire sample.

If the gap fluctuations are smooth, the wire associated with the ZGL will have many 1D subbands occupied. The important point is that two of these subbands will be the Weil states with a linear dispersion
topologically protected from electron backscattering. The Weil states are similar to the edge states in Bernevig-Hughes-Zhang model of the 2D TI edge. In a small ballistic sample the edge state yields a
dissipationless conductance equal to the conductance quantum.

In a large random system with a size exceeding the QW width fluctuation correlation length the ZGLs form a dense network. In this case the 2D conductivity is a combination of the ideal conductance of a
 Weil state and the conductance associated with tunnelling between different ZGLs lying close to each other.

The hopping between the ZGLs is realized via intermediate Dirac states, which lowers the hopping energy (as compared to the total gap) and raises the hopping probability amplitude. In any case one can
use the characteristic hopping length between different ZGLs as some given quantity. Our consideration is based on geometric fractal properties of ZGLs.

We suppose that in our sample  $|\xi-\xi_{c}|<<1$ so that a network of OI and TI domains is formed and the ZGLs cover the entire sample. It is worth noting, however, that in a realistic random network of
channels a low temperature non-zero 2D conductivity may occur only if electrons can tunnel between the adjacent channels, for which a finite channel width is required \cite{mahmoodian2}. One can estimate that :

\begin{equation}
\frac{|w-w_{c}|}{w_{c}}\sim\left(\frac{\hbar v}{a \alpha w_{c}}\right) ^{r}
\end{equation}

where $\alpha =\partial\Delta(w)/\partial w|_{w=w_{c}}$, $\Delta$ is the rms energy gap fluctuation value, $v$ is the electron velocity, $r$ is parameter, which is close  to 0.45 \cite{mahmoodian2}. The
conductivity of the network can be estimated from the equation \cite{mahmoodian2}:

\begin{equation}
\sigma =\frac{e^{2}D}{2\pi h\hbar va}\sim \frac{e^{2}}{h}\left(\frac{\Delta a}{hv}\right)^{p}
\end{equation}

where $D \simeq av (\Delta a/\hbar v)^p$ is the diffusion coefficient , $p$ is a coefficient close to 0.15. We estimate the corresponding HgTe QW parameters
from \cite{mahmoodian2} : $hv = 560 meV$; $\alpha \approx - 8.75 meV/ nm$,$\Delta\approx |\alpha| \sqrt{\overline{w^{2}}-\overline{w}^{2}}\approx 4 meV$, $a=30$ nm.

Combining all parameters we obtain  $\sigma_(n_{w}) = (e^{2}/h) (1\pm 0.1)$, which is smaller than the average experimental value $\sigma_{exp} = (e^{2}/h) (2.5 \pm 1)$.  This disagreement can
be diminished if one takes into account the edge states nonzero width.  This results in new percolation paths and, correspondingly, in higher $\sigma_{nw}$ values and also in a much
stronger $\sigma_{nw}$ dispersion in agreement with the experiment.

The percolation model \cite{mahmoodian2} predicts a non-zero conductivity in a narrow width interval near the critical value $w_{c}$ due to the suppression of backscattering of electrons
propagating alone ZGLs and the growing hopping between adjacent ZGLs as $\overline{w}\rightarrow w_{c}$. It is expected, that the network conductivity vanishes outside of this percolation
threshold. One can estimate the energy and the charge density interval where percolation conductivity occurs. However, in the experiment the total conductivity does not show a crossover
from $\sim e^{2}/h$ values to zero  beyond the percolation threshold, expected from the network model. Instead we observe a smooth conductivity growth with density near the CNP
(Figure 3 and Figure 4b). As we already discussed above, the percolation conductivity is short-circuited by the conductivity of the bulk electrons.

\subsection{The Hall effect in topological channel system}

The percolation model also predicts other characteristic peculiarities in the transport coefficients behavior as well. One of the key effects proving the presence of a network structure in
the system studied is the quenching of the Hall effect at low magnetic field.

The Figures 5a,b  show the  Hall resistivity as a function of gate voltage and magnetic field. Figure \ref{fig:5}b shows a number of representative  $\rho_{xy}(B)$ curves
for $V_g=V_{CNP}$, $V_g<V_{CNP}$ and $V_g>V_{CNP}$. One can see a very narrow $\Delta V_{g}\simeq 0.05 V$ region near the CNP, where $\rho_{xy}\approx0$. Figures 5 a,b show
that the Hall resistivity is flat and close to zero in the interval of magnetic field $-0.1 T < B < 0.1T$. So, according to experiment there is a quenching of the Hall effect at the CNP
in the energy interval corresponding to the density variation \textbf{$\delta n_{s}\sim 5\times10^{9} cm^{-2}$.}

From the theoretical point of view one can treat this situation as follows. A curved 1D ZGL state can be mapped on the QW plane using the coordinate $\textbf{r}(s)$ , where $s$ is the
length along the edge and $\textbf{s}=dr(s)/ds$ is the tangent to the curve. If we take the magnetic field $\textbf{B}=(0,0, B)$ and the corresponding vector
potential $\textbf{A}=\textbf{B}\times \textbf{s}$ then the Hamiltonian of an electron in the 1D ZGL state will be $H=\sigma \nu_{F}(p-e[\textbf{B}\times \textbf{r}(s)]\textbf{s}/c)$ ,
where $\sigma$ is a spin index and $\nu_{F}$ is the Fermi velocity. The gauge transform $U=exp\left(i(e/c)\int[\textbf{B}\times \textbf{r}(s)]ds\right)$  converts the Hamiltonian $H$ to $\sigma p$ .
In this way, as one can see,  the magnetic field  can be effectively eliminated from the Hamiltonian for all open-ended ZDLs which define the character of the low-temperature transport. And this
cancels the Hall current as well. Indeed, the above gauge transform shows that the effect of magnetic field on such a system will be null unless the ZGL are closed or are not single-connected.
But in general the ZGL do not branch out. In fact, the presence of a branching point $(x_{0}, y_{0})$  means a simultaneous fulfillment of three conditions:
$\triangle(x_{0}, y_{0})=0$, $\partial_{x_{0}}(x_{0}, y_{0})=0$  and $\partial_{y_{0}}(x_{0}, y_{0})=0$, which is practically impossible. Hence, the theory predicts a zero Hall
 current in the presence of magnetic field, in agreement with the experimental.

\begin{figure}[ht]
\includegraphics[width=\linewidth]{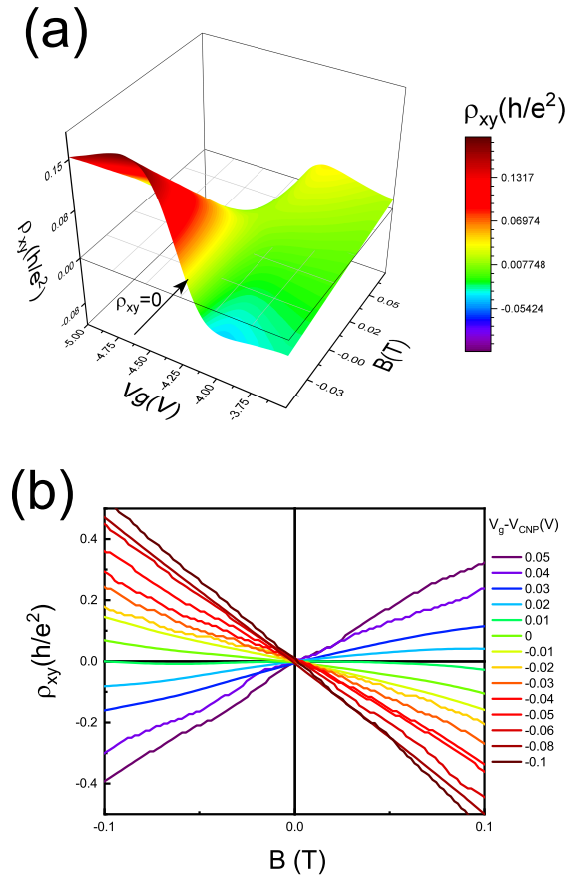}
\caption{(Color online)
(a) The Hall  resistivity as a function of the gate voltage and magnetic field near the CNP.
(b)The  Hall resistivity near the CNP for different values of gate voltage.}
\end{figure}

It worth noting that the bulk contribution to the Hall current will also be absent. Indeed, under these conditions the Fermi energy will be located in a narrow band of localized
states. One can estimate this band width $\delta_{b}$ from the density interval $\delta n_{s} = 5\times10^{9} cm^{-2}$ corresponding to the quenching of the Hall effect (see above).
Using the value of DoS obtained from capacitance measurements \cite{kozlov} yields $\delta_{b}$ = 3 meV. This estimate means that the band width of localized states is close to the
characteristic disorder amplitude in HgTe quantum well.

\section{Methods}
Quantum wells $Cd_{0.65}Hg_{0.35}Te/HgTe/Cd_{0.65}Hg_{0.35}Te$ with (013) surface orientations and a nominal well thickness of (6.3-6.6) nm were prepared by molecular beam
epitaxy.  As shown in the previous publications \cite{gusev}, the use of substrates inclined to the singular orientations facilitates
the growth of more perfect films. Therefore, the growth of alloys is performed predominantly on the substrates with surface orientation [013], which deviates from the singular
orientation by approximately $19^{\circ}$.
Fabrication of ohmic contact to HgTe quantum well is similar to that for other 2D systems, such as GaAs quantum wells, for example: the contacts were formed by the burning-in of
indium directly on the surface of large contact pads. Modulation-doped HgTe/CdHgTe quantum wells are typically grown at 180$^{\rm o}$ C, which is relatively low compared to III-V
compounds. On each contact pad the indium diffuses vertically down, providing ohmic contact to the underlying quantum well, with the contact resistance in the range of 0.1-1 kOhm.
During the AC measurements we made sure that the Y-component of the impedance did not exceed 5\% of the total impedance, which is the indication of good ohmicity of the contacts.
The sample is a Hall bar device with eight voltage probes. The bar has the width $W$ of $50 \mu m$ and three consecutive segments of different lengths $L$ $(100, 250, 100 \mu m )$
(Figure 3, left bottom panel). A dielectric layer was deposited (100 nm of $SiO_{2}$ and 100 nm of $Si_{3}Ni_{4}$) on the sample surface and then covered by a TiAu gate.
The density variation with gate voltage was $1\times10^{11} cm^{-2}V^{-1}$. The magnetotransport measurements were performed in the temperature  $4.2 K$  using a standard four point
circuit with a $1-13 Hz$ ac current of $1-10 nA$ through the sample, which is sufficiently low to avoid overheating effects.

\section{Summary and conclusion}

In conclusion, our results provide an opportunity to take a fresh look at the nature of electron transport in HgTe quantum wells of critical thickness. The description of transport
in this system in terms of percolation through a network of one-dimensional conducting channels makes it possible to study the effects caused by the interplay of topology and localization.
More generally, the study of transport in zero gap HgTe quantum wells can improve our understanding of the disorder induced topological insulator-to-metal transition and  may be important
for a wider class of disordered 2D electron systems, than was considered previously.

\section{Acknowledgment}
G.M. Gusev acknowledges for financial support FAPESP (Brazil) and CNPq (Brazil); Z. D. Kvon, D. A. Kozlov, E.B.Olshanetsky, M.V. Entin and N. N. Mikhailov acknowledge for
financial support  Ministry of Science and Higher Education of the Russian Federation, Grant No. 075-15-2020- 797(13.1902.21.0024).


\section*{References}

\begin{thebibliography}{40}

\bibitem{kane}
Kane, C. L.; Mele, E. J. Z. Topological Order and the
Quantum Spin Hall Effect. {\it Phys.Rev.Lett}. {\bf 2005}, {\it 95}, 146802.

\bibitem{hasan}
Hasan, M.Z.; Kane, C.L. Topological insulator. {\it Rev.Mod.Phys}. {\bf 2010}, {\it
82}, 2045.  Qi X-L.,  Zhang, S-C. Topological insulators and
superconductors. {\it Rev.Mod.Phys.} {\bf 2011}, {\it 83}, 1057.

\bibitem{qi}
Qi, X-L.; Zhang, S-C. The quantum spin Hall effect and topological
insulators. {\it Phys. Today}. {\bf 2010}, {\it 63(1)}, 33 .

\bibitem{moore}
Moore, J. E.; Balents, L. Topological invariants of
time-reversal-invariant band structures. {\it Phys.Rev.B}.
{\bf 2007},{\it 75}, 121306.

\bibitem{moore2}
Moore, J.E. The birth of topological insulators. {\it Nature}.{\bf 2010}, {\it 464}, 194.

\bibitem{konig}
K\"{o}nig, M.; Wiedmann, S.;  Brune, C.;  Roth, A.; Buhmann, H.;
Molenkamp, L.W.;  Qi, X.-L.; Zhang, S.-C. Quantum Spin Hall
Insulator State in HgTe Quantum. {\it Science}. {\bf 2007}, {\it 318}, 766.

\bibitem{buhman}
Buhmann, H. The quantum spin Hall effect. {\it Journal. Appl.Phys.} {\bf 2011}, {\it 109}, 102409.

\bibitem{roth}
Roth, A.; Br$\ddot{u}$ne, C.;  Buhmann, H.;  Molenkamp, L.W.;
Maciejko, J.;  Qi, X.-L.; Zhang, S.-C. Nonlocal Transport in the
Quantum Spin Hall State. {\it Science}.{\bf 2009}, {\it 325}, 294.

\bibitem{gusev}
Gusev, G. M.; Kvon, Z. D.; Shegai, O. A.;  Mikhailov, N. N.; Dvoretsky,
S. A.; Portal, J. C. Transport in disordered two-dimensional
topological insulators. {\it Phys. Rev. B}. {\bf 2011}, {\it 84}, 121302(R).

\bibitem{olshanetsky}
Olshanetsky, E. B.;  Kvon, Z. D.; Gusev, G.M.; Levin, A.D.; Raichev,
O.E.;  Mikhailov, N. N.;  Dvoretsky, S. A. Persistence of a
Two-Dimensional Topological Insulator State in Wide HgTe Quantum
Wells. {\it Phys.Rev.Lett.} {\bf 2015}, {\it 114}, 126802.

\bibitem{rahim}
Rahim, A.;  Levin, A. D.; Gusev, G. M.; Kvon, Z. D.; Olshanetsky, E.
B.;  Mikhailov, N. N.;  Dvoretsky, S. A. Scaling of local and
nonlocal resistances in a 2D topological insulator based on HgTe
quantum well. {\it 2D Materials}. {\bf 2015}, {\it 2}, 044015.

\bibitem{knez}
Knez, I.;  Du, R.-R.; Sullivan, G. Evidence for Helical Edge
Modes in Inverted InAs / GaSb Quantum Wells. {\it Phys. Rev. Lett.} {\bf 2011},
{\it 107}, 136603.

\bibitem{knez2}
Knez, I.;  Rettner, C. T.; Yang, S.-H.;
Parkin, S. S. P.;  Du, L.;  Du, R.-R.; Sullivan, G. Observation of
Edge Transport in the Disordered Regime of Topologically
Insulating InAs/GaSb Quantum Wells. {\it Phys. Rev. Lett.} {\bf 2014}, {\it 112},
026602.

\bibitem{du}
Du, L.; Knez, I.;  Sullivan, G.; Du, R.-R. Robust Helical Edge
Transport in Gated InAs/GaSb Bilayers. {\it Phys. Rev. Lett.}{\bf 2015},
{\it 114}, 096802.

\bibitem{nichele}
Nichele, F.;  Pal, A. N.; Pietsch, P.;  Ihn, T.;  Ensslin, K.;
Charpentier, C. and  Wegscheider W. Insulating State and Giant
Nonlocal Response in an InAs/GaSb Quantum Well in the Quantum Hall
Regime. {\it Phys. Rev. Lett.} {\bf 2014}, {\it 112}, 036802.

\bibitem{suzuki}
Suzuki, K.;  Harada, Y.; Onomitsu, K. and  Muraki K. Gate-controlled
semimetal-topological insulator transition in an InAs/GaSb
heterostructure. {\it Phys. Rev. B.} {\bf 2015}, {\it 91}, 245309.

\bibitem{bernevig}
Bernevig B. A. Toplogical Insulators and Topological
Superconductors. Princeton University Press, 2013.

\bibitem{onoda}
Masaru Onoda; Yshai Avishai and Naoto Nagaosa. Localization in a
Quantum Spin Hall System. {\it Phys. Rev. Lett.} {\bf 2007}, {\it 98}, 076802.

\bibitem{obuse}
Hideaki Obuse; Akira Furusaki; Shinsei Ryu and Christopher Mudry.
Two-dimensional spin-filtered chiral network model for the Z2
quantum spin-Hall effect. {\it Phys. Rev. B.}{\bf 2007}, {\it 76}, 075301.

\bibitem{obuse2}
Hideaki Obuse; Akira Furusaki; Shinsei Ryu and Christopher Mudry.
Boundary criticality at the Anderson transition between a metal
and a quantum spin Hall insulator in two dimensions. {\it Phys. Rev. B.}{\bf 2008},
{\it 78}, 115301.

\bibitem{bondesan}
Bondesan, R.;  Gruzberg, I. A.;  Jacobsen, J. L.;  Obuse, H. and
Saleur H. Exact Exponents for the Spin Quantum Hall Transition in
the Presence of Multiple Edge Channels. {\it Phys. Rev. Lett.} {\bf 2012},
{\it 108}, 126801.

\bibitem{yamakage}
Ai Yamakage; Kentaro Nomura; Ken-Ichiro Imura and Yoshio
Kuramoto. Criticality of the metal-topological insulator
transition driven by disorder. {\it Phys. Rev. B} {\bf 2013}, {\it 87}, 205141.

\bibitem{bhardwaj}
Bhardwaj, S.;  Mkhitaryan, V. V.;  Gruzberg, I. A. Supersymmetry
approach to delocalization transitions in a network model of the
weak-field quantum Hall effect and related models. {\it Phys. Rev. B.}{\bf 2014}, {\it 89}, 235305.

\bibitem{chalker}
Chalker, J. T. and Coddington, P. D. Percolation, quantum tunnelling
and the integer Hall effect. {\it J. Phys. C: Solid State Phys.}{\bf 1988}, {\it
21}, 2665.

\bibitem{li}
Jian Li; Rui-Lin Chu; Jain, J. K. and Shun-Qing Shen. Topological
Anderson Insulator. {\it Phys. Rev. Lett.} {\bf 2009}, {\it  102}, 136806.

\bibitem{mahmoodian}
Mahmoodian, M. M. and  Entin, M. V. Microwave Absorption in 2D
Topological Insulators with a Developed Edge States Network.
{\it Physica Status Solidi b}.{\bf 2019},{\it  256}, 1800652.

\bibitem{mahmoodian2}
Mahmoodian, M. M. and  Entin,  M. V. Conductivity of a
two-dimensional HgTe layer near the critical width: The role of
developed edge states network and random mixture of p-and
n-domains. {\it Phys.Rev. B.}{\bf 2020}, {\it  101}, 125415.

\bibitem{buttner}
Buttner, B.; Liu, C. X.; Tkachov, G.; Novik, E. G.; Brune, C.; Buhmann, H.; Hankiewicz, E. M.;  Recher, P.;  Trauzettel, B.; Zhang, S. C.
and  Molenkamp, L. W. Single valley Dirac fermions in zero-gap HgTe
quantum wells. {\it Nat. Phys.} {\bf 2011}, {\it  7}, 418.

\bibitem{yahniuk}
Yahniuk, I.; Krishtopenko, S. S.; Grabecki, G.; Jouault, B; Consejo,  C.; Desrat, W.;
Majewicz M.;, Kadykov A. M.;, Spirin K. E.;, Gavrilenko, V. I.; Mikhailov, N. N.; Dvoretsky, S. A.;
But, D. B.; Teppe, F.; Wróbel, J.; Cywiński, G.; Kret,  S.; Dietl, T.; Wojciech Knap, W. Magneto-transport in inverted HgTe quantum wells.
{\it npj Quantum Materials} {\bf 2019}, {\it  4}, 13.

\bibitem{kvon}
Kozlov, D. A.;  Kvon, Z. D.;  Mikhailov, N. N. and   Dvoretskii, S.
A. Weak localization of Dirac fermions in HgTe quantum wells. {\it JETP
Lett.} {\bf 2012}, {\it  96}, 730.

\bibitem{kvon2}
Kozlov, D. A.;  Kvon, Z. D.;  Mikhailov, N. N. and   Dvoretskii, S.
A. Quantum Hall effect in a system of gapless Dirac fermions in
HgTe quantum wells. {\it JETP Lett.}{\bf 2014}, {\it  100}, 724.

\bibitem{gusev2}
Gusev, G. M.; Kozlov, D. A.;  Levin, A. D.;  Kvon, Z. D.; Mikhailov, N.
N. and Dvoretsky, S. A. Robust helical edge transport at  $\nu=0$
quantum Hall state. {\it Phys. Rev. B.} {\bf 2017}, {\it  96}, 045304.

\bibitem{davies}
J. Davies, The Physics of Low Dimensional Semiconductors, Cambridge, 1997.

\bibitem{kvon3}
Dobretsova, A. A.;   Kvon, Z. D.;  Braginskii, L. S.;  Entin, M. V.
and Mikhailov, N. N. Mobility of Dirac electrons in HgTe quantum
wells. {\it JETP Lett.} {\bf 2016}, {\it  104}, 388.

\bibitem{efros}
Shklovskii, B. I.;  Efros, A. L. Electronic Properties of Doped
Semiconductors. Springer, Heidelberg, 1984.

\bibitem{reno}
Das Sarma, S.;  Lilly, M. P.;  Hwang, E. H.;  Pfeiffer, L. N.; West,
K.W. and  Reno, J.L. Two-Dimensional Metal-Insulator Transition as
a Percolation Transition in a High-Mobility Electron System. {\it Phys.
Rev. Lett.} {\bf 2005}, {\it  94}, 136401.

\bibitem{kozlov}
Kozlov, D. A. ;  Savchenko, M. L;  Ziegler, J.;  Kvon, Z. D.;
Mikhailov, N. N.;  Dvoretskii, S. A. and  Weiss, D.  Capacitance
spectroscopy of a system of gapless Dirac fermions in a HgTe
quantum well. {\it JETP Lett.}{\bf 2016}, {\it 104}, 859.

\end{thebibliography}
\end{document}